\documentclass[12pt]{iopart}

\usepackage{iopams} 
\usepackage{verbatim}
\usepackage{graphicx}
\usepackage{color}
\begin{document}

\title{Straintronic magnetic tunnel junctions for analog computation: A perspective }

\author{Supriyo Bandyopadhyay}

\address{Department of Electrical and Computer Engineering, Virginia Commonwealth University, Richmond, VA 23284, USA}
\ead{sbandy@vcu.edu}
\vspace{10pt}

\begin{abstract}
The straintronic magnetic tunnel junction (s-MTJ) is an MTJ whose resistance state can be changed {\it continuously} or gradually from high to low  with a gate voltage that generates  strain the magnetostrictive soft layer. This unusual feature, not usually available in MTJs that are switched abruptly with spin transfer torque, spin-orbit torque or voltage-controlled-magnetic-anisotropy, enables many analog applications where the typically low tunneling magneto-resistance ratio of MTJs (on/off ratio of the switch) and the relatively large switching error rate are not serious impediments unlike in digital logic or memory. More importantly, the transfer characteristic of a s-MTJ (conductance versus gate voltage) always sports a {\it linear} region that can be exploited to implement analog arithmetic, vector matrix multiplication and linear synapses in deep learning networks very effectively. In these applications, the s-MTJ is actually superior to the better known memristors and domain wall synapses which do not exhibit the linearity and/or the analog behavior. \footnote{Invited perspective}

\end{abstract}

\medskip


%
%
%
\maketitle
%
%

\section{Introduction: The straintronic magnetic tunnel junction}

\begin{figure}[h!]
\centering
\includegraphics[width=6.5in]{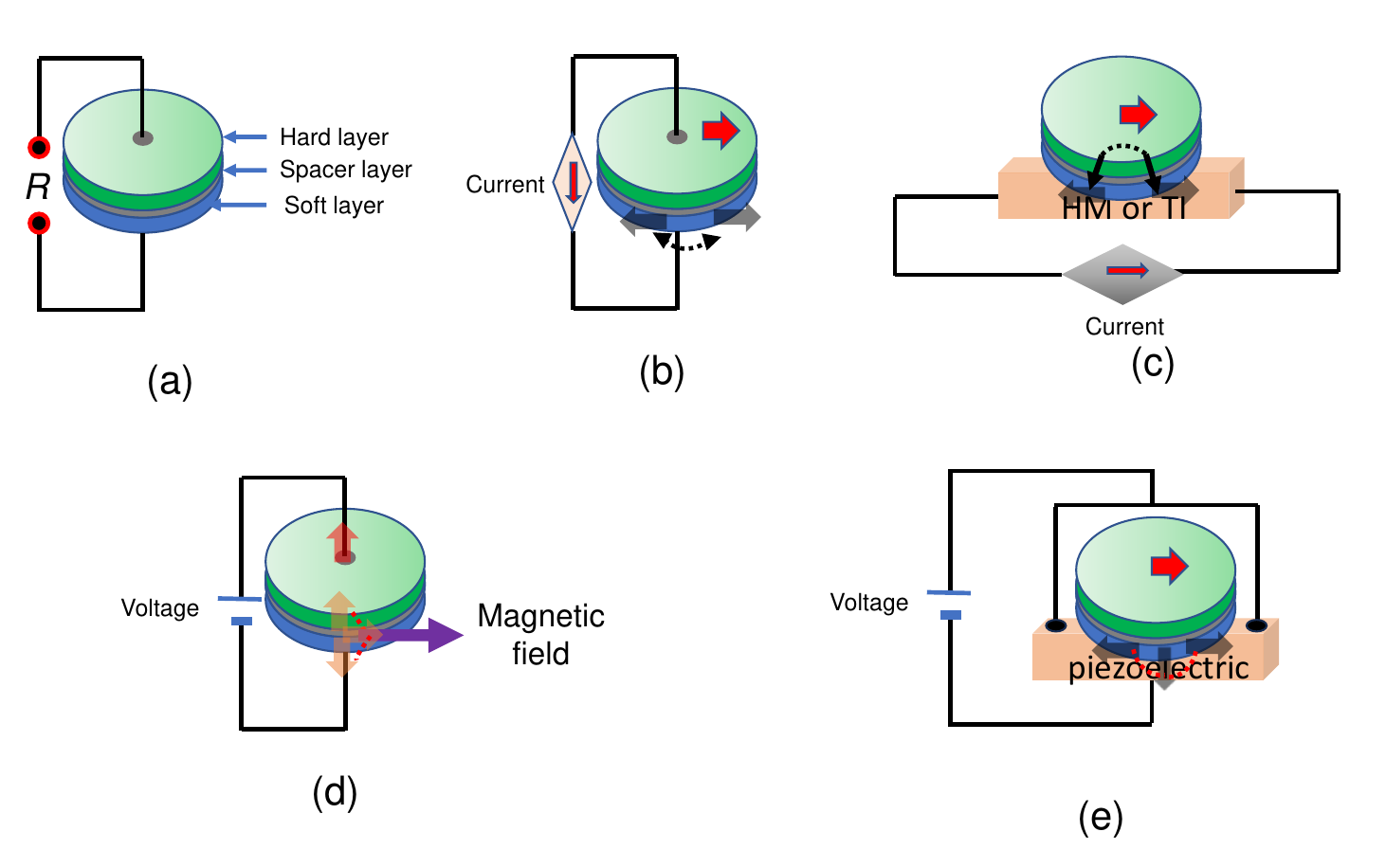}
    \caption{\small (a) A magnetic tunnel junction (MTJ). (b) Switching the MTJ resistance with spin transfer torque (STT). (c) Switching the resistance with spin-orbit torque (SOT). (d) Switching the resistance with voltage-controlled-magnetic-anisotropy (VCMA). (d) Straintronic switching. The diagrams are not to scale.}
\label{fig:MTJ}
\end{figure}

Magnetic tunnel junctions (MTJs) are the quintessential spin-to-charge converters which convert magnetic information into electrical information. They are widely used in magnetic memory and logic as a binary switch. The archetypal MTJ is shown in Fig. \ref{fig:MTJ}(a) and consists of an insulating spacer layer sandwiched between two ferromagnetic layers of different material composition or different thickness. All layers are slightly elliptical in cross-section, so that the magnetization of both ferromagnetic layers has an easy axis along the major axis of the ellipse and hence will tend to align along one of the two directions parallel to the major axis.

One of the ferromagnetic layers has a fixed magnetization (along the major axis) and is referred to as the ``hard layer'', while the other’s magnetization can be changed by an external agent, such as a current or a voltage. It is referred to as the ``soft layer''. The resistance of the MTJ, $R$, measured between the two ferromagnetic layers, is determined by spin-dependent tunneling of electrons between the two ferromagnetic layers, through the insulating spacer layer, and therefore depends on the relative magnetization orientations of the two ferromagnetic layers. The resistance is given approximately by the expression 
\begin{equation}
 R (\theta) = R_P + {{R_{AP} - R_P}\over{2}} [ 1 - cos \theta]   ,
\end{equation}
where $\theta$ is the angle between the magnetizations of the hard and the soft layers, $R_P$ is the resistance when the two magnetizations are mutually parallel and $R_{AP}$ is the resistance when the two magnetizations are mutually antiparallel. Usually, $R_{AP} > R_P$.

The above expression for the resistance shows that one can change it by changing the angle $\theta$ with an external agent. The external agent can be a current passed through the MTJ which will tend to make the magnetizations parallel or antiparallel depending on the polarity of the current via the mechanism of spin transfer torque (STT) \cite{berger,slonczewski}. Another technique is to place the MTJ on a strip of heavy metal (HM) like Pt or $\beta$-Ta (with the soft layer in contact with the strip) and pass a current through the strip which exerts a spin-orbit torque (SOT) on the soft layer’s magnetization to rotate it  \cite{ralph}, making it either parallel or antiparallel to the magnetization of the hard layer and thereby switching the MTJ’s resistance from $R_P$ to $R_{AP}$. or vice versa. The heavy metal can be replaced with a topological insulator (TI) \cite{mellnik} or Weyl semi-metal \cite{liu} which also exerts a spin-orbit torque to rotate the soft layer’s magnetization to effect the switching. 

Yet another technique is to use a voltage instead of a current and it is best suited to ferromagnetic materials whose stable magnetization orientations are out of plane (``up'' and ``down'') owing to perpendicular magnetic anisotropy. The voltage changes the anisotropy of the soft layer from perpendicular to in-plane and hence rotates the magnetization by 90$^{\circ}$, bringing it from out-of-plane to in-plane. An in-plane magnetic field then causes the magnetization to precess further around it and the voltage pulse is shaped such that it reaches zero when the precession has caused another 90$^{\circ}$ rotation to complete a full 180$^{\circ}$ rotation of the magnetization. This is referred to as voltage controlled magnetic anisotropy (VCMA) based switching \cite{pedram}.

The fourth methodology is relatively uncommon and will work only when the soft layer is made of a magnetostrictive material that is placed in elastic contact with a piezoelectric layer. A voltage applied across the piezoelectric generates biaxial strain underneath the MTJ \cite{cui} and strains the soft layer whose magnetization rotates from the major axis toward the minor axis depending on the sign of the strain (compressive or tensile) \cite{pra}. If the strain is withdrawn as soon as the magnetization has rotated through 90$^{\circ}$ in the soft layer's plane, then a residual torque on the magnetization associated with the out-of-plane excursion of the magnetization during rotation will result in another 90$^{\circ}$ rotation in-plane to complete a full 180$^{\circ}$ rotation \cite{kuntal}. There are other (more complicated) ways of flipping the magnetization with strain  which do not require precise timing of the voltage pulse (similar to VCMA), but they require applying voltage pulses sequentially using different pairs of gates \cite{ayan}. The modality of switching magnetization with strain (where the strain can be generated electrically with the aid of a piezolectric layer placed underneath the soft layer) is referred to as ``straintronics'' \cite{nanotechnology}. 

The two voltage-controlled mechanisms –- VCMA and straintronics –- are more energy-efficient than the two current controlled mechanisms (STT and SOT), but also much more error-prone. There is seemingly always a trade-off between energy cost and reliability in binary switching which is almost universal \cite{mdpi}. The four magnetization flipping mechanisms that were discussed are shown in Figs. \ref{fig:MTJ}(b) - (e).

\begin{figure}[!b]
\centering
\includegraphics[width=5.3in]{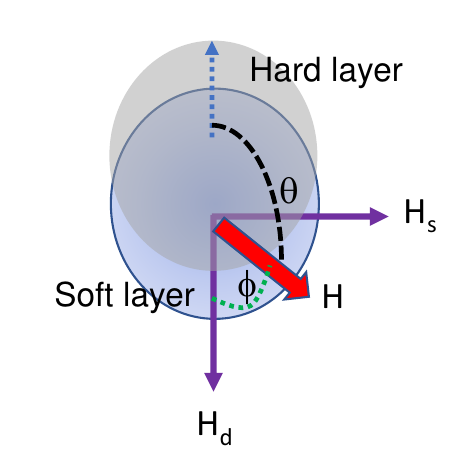}
    \caption{\small The net effective magnetic field experienced by the soft layer of a straintronic magnetic tunnel junction in the presence of: (1) strain of the appropriate sign and (2) dipole coupling with the hard layer. ${\bf H}_s$ is the effective magnetic field due to strain and ${\bf H}_d$ is the effective magnetic field due to dipole coupling with the hard layer. By varying ${\bf H}_s$ continuously with a gate voltage, one can vary $\phi$ and hence the resistance of the s-MTJ continuously between $R_P$ and $R_{AP}$.}
\label{fig:resultant}
\end{figure}

One very important distinction between straintronics and the other three switching mechanisms is that the other three usually can rotate the magnetization through 180$^{\circ}$ and not any arbitrary angle. This will allow the MTJ resistance to assume only two values, $R_P$ and $R_{AP}$. Hence those three mechanisms can implement a binary switch good for digital applications. Straintronics can also implement a binary switch, but additionally  it is possible to rotate the magnetization of the soft layer through any arbitrary angle $\phi$ between 0$^{\circ}$ and 90$^{\circ}$ and orient the magnetization anywhere between the easy and hard axes, i.e, anywhere between the major and minor axes \cite{matrix-paper} as long as the voltage generating the strain is kept on. Any intermediate orientation is very stable against thermal noise since the potential well that forms around that state is very deep (several tens of kT) \cite{arxiv}. This will allow the MTJ resistance to assume arbitrary values between $R_P$ and $R_{AP}$, which lends itself to {\it analog applications} that are far less explored than digital applications of MTJs and yet can be more rewarding since they are usually more error-tolerant.

The above feature that allows a straintronic MTJ's resistance to assume any arbitrary value between $R_P$ and $R_{AP}$ requires some amount of dipole coupling between the hard and the soft layers of the MTJ. Dipole coupling acts like an effective magnetic field ${\bf H}_d$ along the easy (major) axis of the soft layer whose direction is antiparallel to the magnetization of the hard layer \cite{matrix-paper}. The strain, on the other hand, acts like an effective magnetic field ${\bf H}_s$ along the hard axis (minor axis) of the elliptical soft layer. The resultant effective magnetic field ${\bf H}$ will point along the vector sum of the two fields as shown in Fig. \ref{fig:resultant}. By varying the strain continuously with an analog voltage, we can continuously vary ${\bf H}_s$  and hence change the direction of ${\bf H}$ (or the angle $\phi$). Since the magnetization will ultimately align along ${\bf H}$, we can rotate the magnetization of the soft layer through any angle $\phi$ between 0$^{\circ}$ and 90$^{\circ}$, and hence change the resistance of the MTJ continuously in an analog fashion.

\begin{figure}[!t]
\centering
\includegraphics[width=6in]{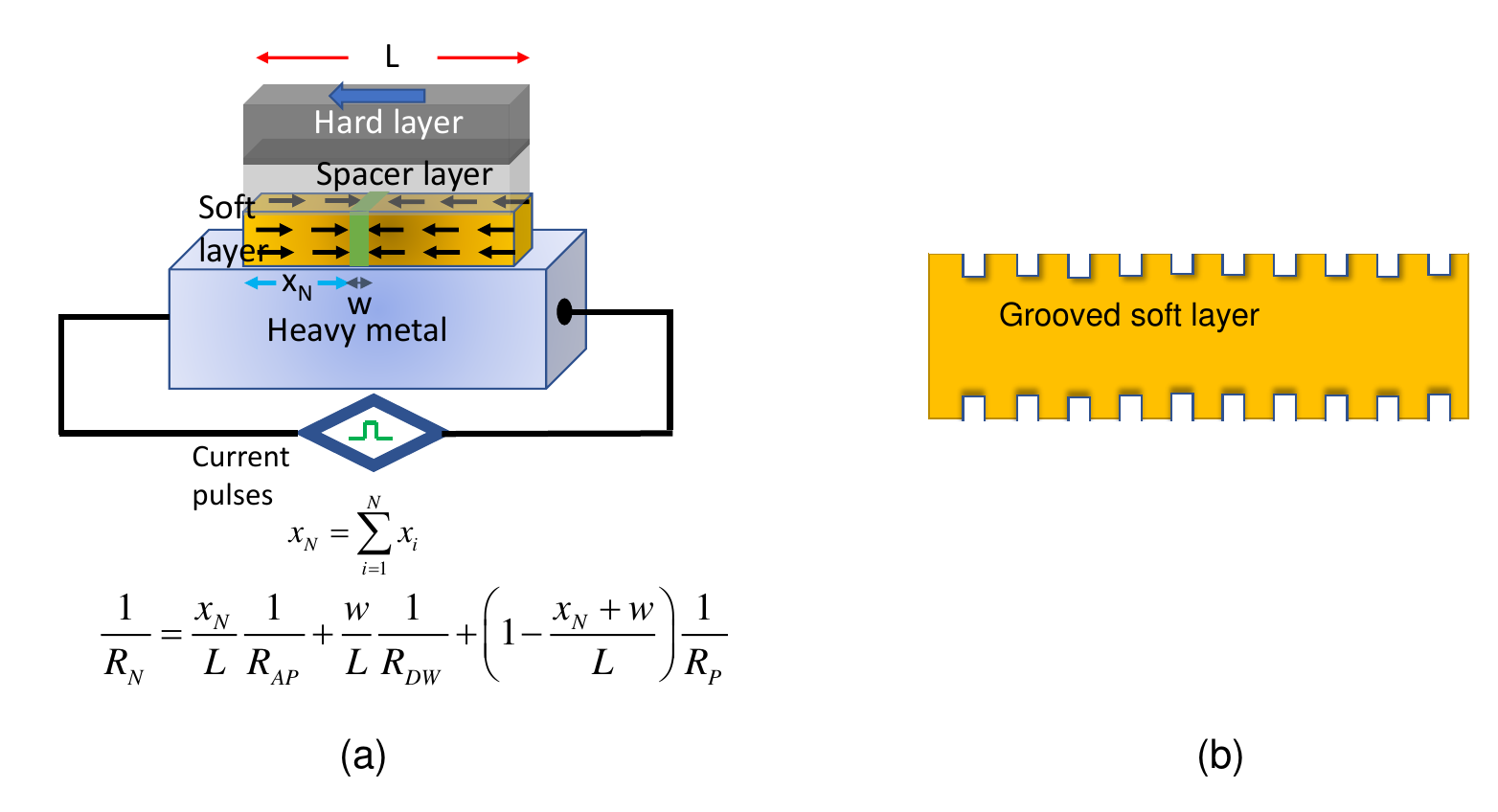}
    \caption{\small (a) A domain wall synapse consisting of a magnetic tunnel junction placed over a heavy metal layer. Successive current pulses passed through the heavy metal exerts successive spin-orbit torques on the soft layer that is in contact with the heavy metal. These pulses move the domain wall in steps thereby changing the fraction of the soft layer whose magnetization is parallel (antiparallel) to that of the hard layer. The resistance of the MTJ is the parallel combination of three resistances consisting of the fraction where the magnetizations of the hard and soft layer are parallel, the fraction where they are antiparallel and the remaining fraction involving the domain wall. (b) A grooved soft layer where the domain wall position is stabilized against ``creep'' which is back-and-forth motion of the domain wall due to thermal noise. }
\label{fig:domain-wall}
\end{figure}

There are other {\it quasi-continuous} ways of varying an MTJ’s resistance that do not involve straintronics. One is by propagating a domain wall through a rectangular soft layer in steps \cite{grollier,incorvia1,incorvia2}. Successive current pulses applied through a heavy-metal or topological insulator underlayer (see Fig. \ref{fig:MTJ} (c)) will generate successive spin orbit torque pulses on the soft layer and move a domain wall along the length of the soft layer in steps of $x_i$. This changes the fraction of the soft layer whose magnetization is parallel (antiparallel) to that of the hard layer {\it in steps} and hence changes the resistance of the MTJ in steps (see Fig. \ref{fig:domain-wall}(a); $R_N$ is the resistance of the MTJ after $N$ current pulses). This, of course, does not enable truly continuous change (since it is in discrete steps), but allows the MTJ resistance to assume (discrete) intermediate values between $R_P$ and $R_{AP}$. However, such an approach is easier said than done. Once the domain wall has moved to a location after the current pulse subsides, it does not necessarily stay there, but can move backwards or forwards owing to thermal noise, which makes the MTJ resistance unstable. There are some proposed approaches to counter this effect, such as by making the soft layer grooved as shown in Fig. \ref{fig:domain-wall}(b), but it is not 100\% reliable. The shape of the grooves, the depth of the grooves and the separation between neighboring grooves –- all have to be carefully designed and controlled to minimize undesired domain wall movement or ``creep'' due to thermal noise. Even then, stability of the MTJ resistance could be elusive.

There are many other ways of changing an MTJ’s resistance in a quasi-analog fashion, such as by gradually changing the magnetization of a ferromagnetic soft layer exchange coupled to an antiferromagnetic layer \cite{fukami} via pulsed current flow through the antiferromagnetic layer, or by switching the grains of a granular soft layer one after another \cite{wang}, or by nucleating an increased number of skyrmions in the soft layer \cite{song}. These are difficult-to-control techniques and they are only quasi-continuous, not completely continuous. In contrast, the method involving straintronics to change an MTJ’s resistance can be truly continuous and has other desirable properties such as: (1) any intermediate state between the 0$^{\circ}$ and 90$^{\circ}$ (0$^{\circ}$ $\le$ $\phi$ $\le$ 90$^{\circ}$)  is extremely {\it stable} against thermal noise \cite{arxiv}, and (2) the conductance of the MTJ, $G_{s-MTJ}$, can be varied {\it linearly} with a gate voltage $V_G$. The linearity is extremely valuable since it is generally elusive. Many applications in analog computing will require and benefit from this linearity.

\section{Transfer characteristic of a straintronic MTJ}

\begin{figure}[!h]
\centering
\includegraphics[width=6in]{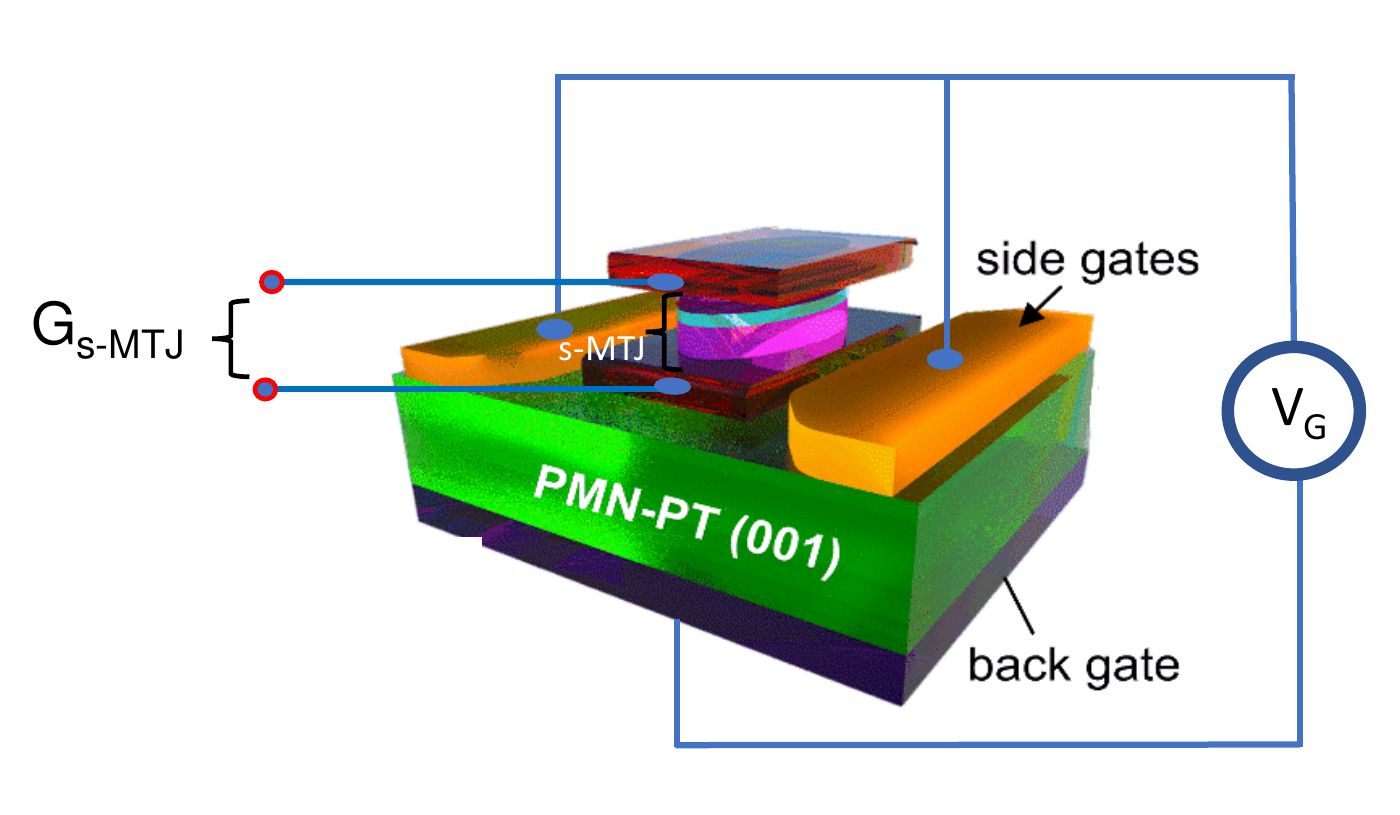}
    \caption{\small The structure of a straintronic magnetic tunnel junction reproduced from \cite{jianping} with permission of the American Institute of Physics.}
\label{fig:s-MTJ}
\end{figure}

The resistance of a straintronic MTJ (s-MTJ) can be changed with strain generated by a gate voltage \cite{jianping} in the configuration shown in Fig. \ref{fig:s-MTJ}. A gate voltage is applied between the two side gates (shorted together) and the back gate \cite{jianping}. The strain rotates the magnetization of the soft layer in contact with the piezoelectric (e.g., PMN-PT) through an angle $\phi$ and that changes the resistance of the s-MTJ according to Equation (1). Here, $\phi = 180^{\circ} - \theta$.  The transfer characteristic is the conductance of the s-MTJ ($G_{s-MTJ}$ ) as a function of the gate voltage $V_G$.

Ref. \cite{matrix-paper} provided an analytical proof that over a range of gate voltage, the transfer characteristic will be linear so that the s-MTJ conductance will be expressed as 
\begin{equation}
    G_{s-MTJ} = G_{AP} + \kappa \left ( V_G - \delta \right ).
    \label{eq:linear}
\end{equation}
where $\kappa$ and $\delta$ are constants and $G_{AP} = 1/R_{AP}$. The quantity  $\kappa = -1/\left (2 R_{AP} \Gamma \right )$, $\delta = \Gamma - \nu$, $\Gamma = \left ( \mu_0 M_s |{\bf H}_d | t \right )/\left ( 3 \lambda_s Y d_{33} \right)$ and $\nu = \left [ M_s \left (N_{min} - N_{maj} \right )/|{\bf H}_d | \right ] \Gamma$, where $\mu_0$ is the permeability of free space, $M_s$ is the saturation magnetization of the soft layer, $t$ is the thickness of the piezoelectric layer, $\lambda_s$ is the saturation magnetostriction of the soft layer, $Y$ is the Young’s modulus of the soft layer, $d_{33}$ is the diagonal component of the piezoelectric coefficient tensor of the piezoelectric layer, while $N_{min}$ and $N_{maj}$ are the demagnetization factors along the minor and major axis of the elliptical soft layer, which depend on the dimensions of the soft layer \cite{chikazumi}. This linear relation was also verified by stochastic Landau-Lifshitz Gilbert simulations at room temperature in \cite{matrix-paper}. 

\begin{figure}[!h]
\centering
\includegraphics[width=6in]{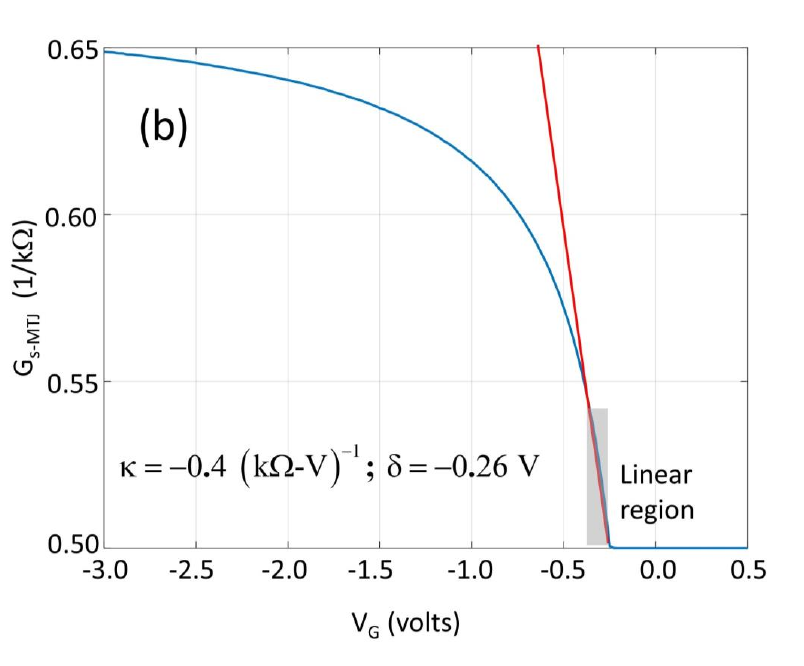}
    \caption{\small The transfer characteristic of a straintronic magnetic tunnel junction computed with the use of stochastic Landau-Lifshitz-Gilbert simulation of the soft layer's magnetodynamics in the presence of 
    gate-voltage-generated strain at room temperature. This figure is reproduced from \cite{matrix-paper} with the permission of the Institute of Electrical and Electronics Engineers. The parameters for the soft layer were major axis = 800 nm, minor axis = 700 nm, thickness = 2.2 nm, saturation magnetization $M_s$ = 8.5$\times$10$^5$ A/m, dipole coupling field $H_d$ = 1000 Oe, Gilbert damping constant = 0.1, saturation magnetostriction $\lambda_s$ = 600 ppm, Young's modulus $Y$ = 120 GPa, piezoelectric coefficient $d_{33}$ = 1.5$\times$10$^{-9}$ C/N and the piezoelectric layer thickness $t$ = 1 $\mu$m. The value of $R_{AP}$ was assumed to be 2 k$\Omega$ and the value of $R_P$ was 1 k$\Omega$.}
\label{fig:transfer}
\end{figure}

Fig. \ref{fig:transfer} shows the computed transfer characteristic of a straintronic MTJ reproduced from \cite{matrix-paper}. There is a range of gate voltage $V_G$ over which the s-MTJ conductance $G_{s-MTJ}$ is a linear function of the gate voltage. This linear region has a small extent in Fig. \ref{fig:transfer}, but it can be extended by increasing the antiparallel resistance $R_{AP}$, perhaps by making the MJT cross-section smaller or the spacer layer thickness larger.

\section{Analog arithmetic}

\begin{figure}[!h]
\centering
\includegraphics[width=6.5in]{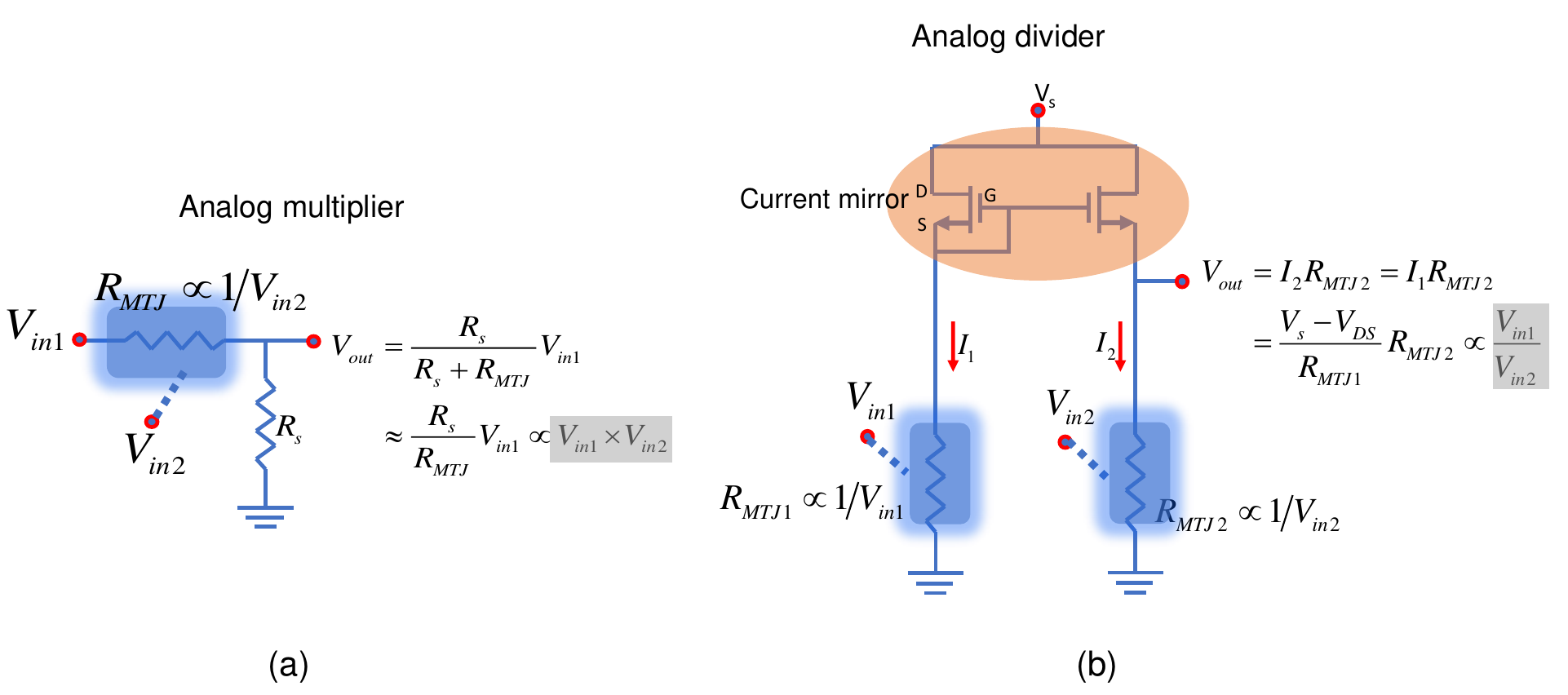}
    \caption{\small (a) An analog multiplier and (b) an analog divider realized with a voltage-dependent resistor whose resistance is inversely proportional to a voltage applied to it.}
\label{fig:arithmetic}
\end{figure}

One intriguing application of the linear transfer characteristic of a s-MTJ is in analog arithmetic. In Fig. \ref{fig:arithmetic}, we show simple circuit representations of an analog multiplier and an analog divider implemented with a voltage-dependent resistor whose conductance is proportional to an external voltage. This voltage-dependent resistor can be realized with a straintronic MTJ operating in the linear region of the transfer characteristic where Equation (\ref{eq:linear}) holds. 

\begin{figure}[!h]
\centering
\includegraphics[width=6.5in]{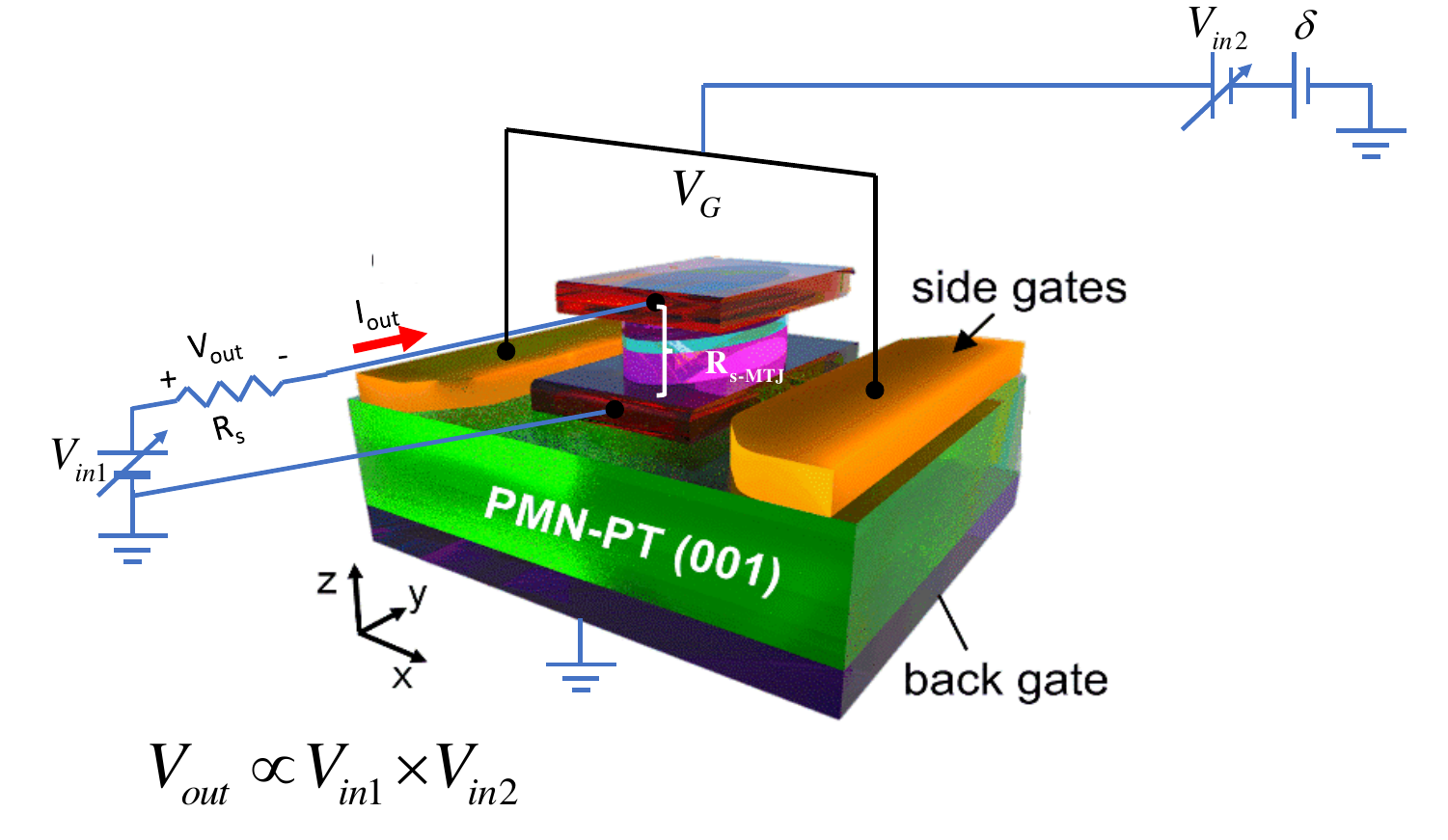}
    \caption{\small (a) An analog multiplier implemented with a straintronic magnetic tunnel junction. The magnetic tunnel junction figure is reproduced from \cite{jianping} with permission of the American Institute of Physics.}
\label{fig:multiplier}
\end{figure}

The actual multiplier implementation is shown in Fig. \ref{fig:multiplier} where the output voltage $V_{out}$ is proportional to the product of the two input voltages $V_{in1}$ and $V_{in2}$, which are the multiplier and the multiplicand. It is obvious from that figure that $V_{in2} = V_G - \delta$. Using this relation in Equation (\ref{eq:linear}), we get $G_{s-MTJ} = G_{AP} + \kappa V_{in2} \approx \kappa V_{in2}$ since $G_{AP}$ is typically small. From Fig. \ref{fig:multiplier}, the voltage over the resistor $R_s$ is $V_{out} = {{R_s}\over{R_s + R_{s-MTJ}}}V_{in1} \approx {{R_s}\over{R_{s-MTJ}}} V_{in1} = G_{s-MTJ}R_s V_{in1} = \kappa R_s \left ( V_{in1} \times V_{in2} \right )$ (since $R_s \ll R_{s-MTJ}$). Thus, the outout voltage $V_{out}$ is proportional to the product of the two voltages $V_{in1}$ and $V_{in2}$. This implements the {\it analog multiplier}. A similar approach can be taken to implement the divider. Thus, the straintronic MTJ lends itself to elegant ways to implement analog arithmetic.

\section{Analog vector matrix multiplier}

\begin{figure}[!h]
\centering
\includegraphics[width=6.5in]{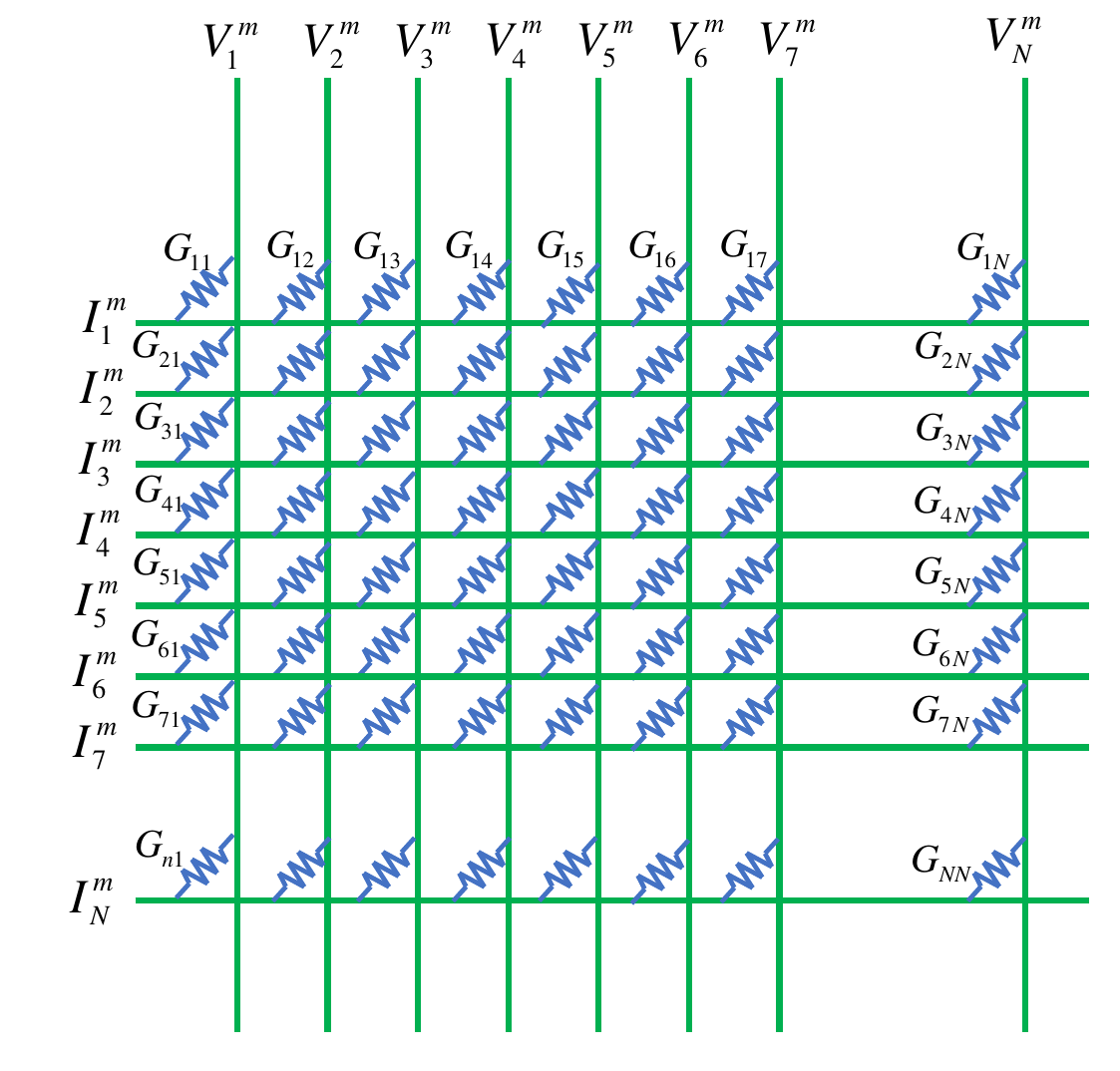}
    \caption{\small The traditional crossbar implementation of a vector matrix multiplier based on Ohm's law and Kirchoff's current law.}
\label{fig:crossbar}
\end{figure}

Vector matrix multiplication is the most important mathematical operation in machine learning and forms the backbone of deep learning networks \cite{Hammerly,ansari}. The well-established approach to implementing a vector matrix multiplier is with a crossbar that leverages Ohm’s law and Kirchoff’s current law to perform matrix multiplication. The basic idea is shown in Fig. \ref{fig:crossbar}. 

Consider the matrix multiplication operation $[{\bf c}] = [{\bf a}] \times [{\bf b}]$ where an element of the product matrix is given by $c_{ij} = \sum_m a_{im} b_{mj}$. The crossbar array shown in Fig. \ref{fig:crossbar} can produce one column of the product matrix, say the $m$-th column $c_i^m (i = 1 \cdot \cdot N)$. The current  $I_i^m$ in the $i$-th horizontal line is given by $I_i^m = \sum_j G_{ij}V_j^m$   based on Ohm’s law and Kirchoff’s current law ($V_j^m$ is the voltage applied to the $j$-th vertical line). The conductance matrix $G_{ij}$ encodes the matrix [{\bf a}]  and the voltage matrix $V_{ji}$ encodes the matrix [{\bf b}]. When both [{\bf a}]  and  [{\bf b}] are $N \times N$ matrices, we will need $N$ such crossbars. Since each crossbar has $N^2$ conductors, we will need a total to $N^3$ conductors (s-MTJs) to perform the matrix multiplication operation. The MTJ-based implementation in \cite{matrix-paper} has an advantge over this popular implementation in that it requires only 2$N^2$ MTJs to perform the same task (since only two MTJs are required to multiply one row with one column instead of $N$ MTJs), but it is a quasi-analog matrix multiplier, albeit non-binary. 

The most important consideration in the normal crossbar approach is what device is used to implement the conductors. Normally, the elements of the multiplier matrix and the multiplicand matrix will be encoded in the {\it same physical quantity}, such as voltage. In that case, we will need the chosen device to have voltage-dependent conductance and the voltage dependence should be linear so that the conductance is proportional to the voltage. {\it This linearity is crucial}. 

Many devices that have been proposed to implement the conductors in crossbar matrix multipliers lack this linearity, such as memristors \cite{memristor1, memristor2} whose conductance usually does not depend linearly on a voltage applied to it to tune the conductance. The same is true of CMOS based conductors \cite{CMOS}. Lately, grooved domain wall synapses, working on the principle in Fig. \ref{fig:domain-wall} have been proposed to implement the conductors \cite{debanjan}. Their conductance can be varied by varying the current density injected into the heavy metal layer to move the domain wall. The conductance is determined by the domain wall position and that can vary quasi-linearly with the current density (and hence with an applied voltage) in a properly designed grooved structure \cite{matrix-paper}.  However, their conductance cannot be varied continuously in a truly analog fashion since the domain wall position, which determines the conductance of the MTJ, can be stabilized only at discrete locations by the grooves and hence the conductance can take only {\it discrete values} \cite{atul}. In contrast, the straintronic MTJ does not suffer from these drawbacks. The conductance can be varied continuously {\it and} linearly with the gate voltage. 

\section{Linear synapses for learning and inference in artificial neural networks}

The straintronic MTJ behaves as a truly ``linear synapse'' whose ``weight'' can be varied continuously and in a linear fashion. That of course makes it attractive for matrix multiplication, but there are also other applications, such as on-chip learning in analog hardware neural networks, where these two properties –- analog behavior and linearity –- are extremely valuable. The linearity also leads to ``symmetry'', whereby equal decrements and increments of gate voltage magnitude will lead to equal increments and decrements of the conductance. In that case, positive weight update of the synapse (long term potentiation) and negative weight update (long term depression) will require voltage pulses of the same magnitude and opposite polarity. This is very desirable for classification tasks in neural networks \cite{debanjan}, where the accuracy of classification benefits from the linearity \cite{debanjan1}. 

\section{Extending the linear region in the transfer characteristic of a s-MTJ}

The transfer characteristic of the straintronic MTJ is given in Equation (\ref{eq:linear})  and it is obvious that we may be able to extend the range of gate voltage where the linear behavior is manisfested if we reduce the slope $\kappa$. This quantity is given by $\kappa = -1/\left (2 R_{AP} \Gamma \right ) = - \left (3 \lambda_s Y d_{33} \right )/\left (2 R_{AP} \mu_0 M_s |{\bf H}_d| t \right )$. Reducing $\lambda_s$ or $d_{33}$, or increasing the piezoelectric layer thickness $t$ is not advisable since that will increase the energy dissipation associated with modulation of the conductance with gate voltage or strain. Hence the best option seems to be to increase the antiparallel resistance $R_{AP}$ by making the MTJ cross-section smaller and/or making the spacer layer thickness larger.

\section{Conclusion}

Here, we have shown that a strain switched magnetic tunnel junction has a unique transfer characteristic with a linear region that can be gainfully exploited for analog computing such as analog arithmetic, analog vector matrix multiplication and analog artificial neural networks. The type of analog computing applications discussed here requires the following features; (1) linearity of control whereby the synaptic weight (s-MTJ conductance) can be varied linearly with a gate voltage over a finite range, (2) stability of the conductance state against external perturbations such as thermal noise, (3) symmetry of the weight update whereby positive and negative weight updates of the same magnitude require same gate voltage but of opposite polarities. The s-MTJ satisfies all these requirements and is hence a superior choice for these analog applications over popular ones like memristors and domain wall synapses.

\section*{Acknowledgment}

This work is supported by the US Air Force Office of Scientific Research under grant FA865123CA023. The author is indebted to Prof. Csaba Andras Moritz of the University of Massachusetts at Amherst for suggesting the resistive implementation of the multiplier and divider as shown in Fig. \ref{fig:arithmetic}.


\bigskip

\section*{References}

\begin{thebibliography}{10}

\bibitem{berger}
Berger L 1996 Emission of spin waves by a magnetic multilayer traversed by a current {\it Phys. Rev. B}, {\bf 54}, 9353.

\bibitem{slonczewski}
Slonczewski J C 1996 Current-driven excitation of magnetic multilayers {\it J. Magn. Magn. Mater.}, {\bf 159}, Li-L7.



\bibitem{ralph}
Liu L, Pai C F, Li Y, Tseng H W, Ralph D C and Buhrman R A 2012 Spin-torque switching with the giant spin Hall effect of tantalum {\it Science} {\bf 336} 555.

\bibitem{mellnik}
Mellnik A R, et al. 2014 Spin-transfer torque generated by a topological insulator {\it Nature}, {\bf 511}, 449-451. 

\bibitem{liu}
MacNeill D, Stiehl G M, Guimaraes M H D, Buhrman R A, Park J and Ralph D C 2017
Control of spin–orbit torques through crystal symmetry in WTe2/ferromagnet bilayers {\it Nat. Phys.}, {\bf 13}, 300-305.

\bibitem{pedram}
Khalili P K and Wang K L 2012 Voltage-controlled magnetic anisotropy in spintronic devices {\it SPIN}, {\bf 2}, 1240002.

\bibitem{cui}
Cui J Z, Liang C Y, Paisley E A, Sepulveda A, Ihlefeld J F, Carman G P and Lynch C S 2015 Generation of localized strain in a thin film piezoelectric to control individual magnetoelectric heterostructures {\it Appl. Phys. Lett.} {\bf 107}, 092903.

\bibitem{pra}
Bandyopadhyay S, Atulasimha J and  Barman A 2021 Magnetic straintronics: Manipulating the magnetization of magnetostrictive nanomagnets with strain for energy-efficient applications {\it Appl. Phys. Rev.}, {\bf 8}, 041323.

\bibitem{kuntal}
Roy K, Bandyopadhyay S and Atulasimha J 2013 Binary switching in a symmetric potential landscape {\it Sci. Rep.}, {\bf 3}, 3038.

\bibitem{ayan}
Biswas A K, Ahmad H, Atulasimha J and  Bandyopadhyay S 2017 Experimental demonstration of complete 180$^{\circ}$ reversal of magnetization in isolated Co nanomagnets on a PMN–PT substrate with voltage generated strain {\it Nano Lett.}, {\bf 17}, 3478-3484.

\bibitem{nanotechnology}
D'Souza N, et al. 2018 
Energy-efficient switching of nanomagnets for computing: Straintronics and other methodologies {\it Nanotechnology}, {\bf 29}, 442001.

\bibitem{mdpi}
Rahman R and Bandyopadhyay S 2021
The cost of energy-efficiency in digital hardware: The trade-off between energy dissipation, energy–delay product and reliability in electronic, magnetic and optical binary switches {\it Appl. Sci.}, {\bf 11}, 5590.

\bibitem{matrix-paper}
Rahman R and Bandyopadhyay S 2022
A nonvolatile all-spin nonbinary matrix multiplier: An efficient hardware accelerator for machine learning {\it IEEE Trans. Elec. Dev.}, {\bf 69}, 7120-7127.

\bibitem{arxiv}
Rahman R and Bandyopadhyay S 2022
A nonvolatile all-spin nonbinary matrix multiplier: An efficient hardware accelerator for machine learning
https://doi.org/10.48550/arXiv.2211.06490. [This preprint has additional material compared to the previous reference.]

\bibitem{grollier} 
Lequeux S, Sampaio J, Cros V, Yakushiji K, Fukushima A, Matsumoto R, Kubota H,  Yuasa S and  Grollier J 2016
A magnetic synapse: multilevel spin-torque memristor with perpendicular anisotropy {\it Sci. Rep.}, {\bf 6} 31510.

\bibitem{incorvia1} 
Liu S, Xiao T P, Cui C,  Incorvia J A C, Bennett C H and Marinella M J 2021
A domain wall-magnetic tunnel junction artificial synapse with notched geometry for accurate and efficient training of deep neural networks {\it Appl. Phys. Lett.}, {\bf 118}, 202405.


\bibitem{incorvia2}
Alamdar M, et al. 2021 Domain wall-magnetic tunnel junction spin–orbit torque devices and circuits for in-memory computing {\it Appl. Phys. Lett.}, {\bf 118}, 112401.

\bibitem{fukami}
Fukami S, Zhang C, DuttaGupta S, Kurenkov A and Ohno H 2016 
Magnetization switching by spin–orbit torque in an antiferromagnet–ferromagnet bilayer system {\it Nat. Mater.}, {\bf 15}, 535-541.

\bibitem{wang}
Wang J, et al. 2019 Impact of carbon segregant on microstructure and magnetic properties of FePt-C nanogranular films on MgO (001) substrate {\it Acta Mater.}. {\bf 166}, 413.


\bibitem{song}
Song K M, et al. 2020 Skyrmion-based artificial synapses for neuromorphic computing {\it Nat. Elec.}, {\bf 3}, 148.

\bibitem{jianping}
Zhao Z, Jamali M, D'Souza N, Zhang D, Bandyopadhyay S, Atulasimha J and Wang J P 2016 Giant voltage manipulation of MgO-based magnetic tunnel junctions via localized anisotropic strain: A potential pathway to ultra-energy-efficient memory technology {\it Appl. Phys. Lett.}, {\bf 109}, 092403.

\bibitem{chikazumi}
Chikazumi S 1964 {\it Physics of Magnetism} (New York: Wiley).

\bibitem{Hammerly}
Hamerly R, Bernstein L, Sludds A, Soljačić M and Englund D 2019 Large-scale optical neural networks based on photoelectric multiplication {\it Phys. Rev. X}, {\bf 9}, 021032.

\bibitem{ansari}
Ansari M A et al. 2019 Improving the accuracy and hardware efficiency of neural networks using approximate multipliers
{\it IEEE Trans. Very Large Scale Int. Syst.}, {\bf 28}, 317-328.

\bibitem{memristor1}
P\'erez-Avila A J, P\'erez E, Rold\'an J B, Wenger C and Jim\'enez-Molinos F 2021 
Multilevel memristor based matrix-vector
multiplication: Influence of the discretization
method {\it 2021 13th Spanish Conference on Electron Devices (CDE)}. DOI: 10.1109/CDE52135.2021.9455724.

\bibitem{memristor2}
Mahmoodi M R, Vincent A F, Nili H and  Strukov D B 2020 Intrinsic bounds for computing precision in
memristor-based vector-by-matrix multipliers {\it IEEE Trans. Nanotechnol.}, {\bf 19}, 429-435.

\bibitem{CMOS}
Amirsoleimani A, et al. 2020 In-memory vector-matrix multiplication in monolithic
complementary metal–oxide–semiconductor-memristor
integrated circuits: Design choices, challenges and
perspectives {\it Adv. Intell. Syst.}, {\bf 2}, 2000115.

\bibitem{debanjan}
Yadav R S, Gupta P, Holla A, Khan K I A, Muduli P K
and Bhowmik D 2023 
Demonstration of synaptic behavior in a 
heavy metal ferromagnetic metal oxide heterostructure based spintronic
device for on-chip learning in crossbar-array-based neural networks
{\it ACS Appl. Electon. Mater.}, {\bf 5}, 484-497.

\bibitem{atul}
Dhull S, Al Misba W, Nisar A, Atulasimha J and Kaushik B K 2024 Quantized magnetic domain wall synapse for efficient deep neural networks {\it IEEE Trans. Neural Networks and Learning Systems (Early Access)}, DOI: 10.1109/TNNLS.2024.3369969.

\bibitem{debanjan1}
Desai V B,  Kaushik D, Sharda J and Bhowmik D 2022 On-chip learning of a domain-wall-synapse-crossbar-array-based convolutional neural network {\it Neuromorphic Computing and Engineering}, {\bf 2}, 024006.


\end{thebibliography}
\end{document}